\documentclass[aps,prl,twocolumn,amsmath,showpacs]{revtex4}
\usepackage{graphicx}

\newcommand{\fv}{{\bf f}}
\newcommand{\kv}{{\bf k}}
\newcommand{\nv}{{\bf n}}
\newcommand{\qv}{{\bf q}}
\newcommand{\rv}{{\bf r}}
\newcommand{\sv}{{\bf s}}

\newcommand{\dk}{\Delta{\bf k}}

\newcommand{\n}[1]{{\bf n}^{(#1)}}
\newcommand{\K}[1]{{\bf K}^{(#1)}}
\newcommand{\A}[1]{{\bf A}^{(#1)}}
\newcommand{\q}[1]{{\bf q}^{(#1)}}
\renewcommand{\a}[1]{{\bf a}^{(#1)}}
\renewcommand{\b}[1]{{\bf b}^{(#1)}}
\renewcommand{\k}[1]{\Delta{\bf k}^{(#1)}}
\newcommand{\sinc}{{\rm sinc}}

\begin{document}

\title{Photonic quasicrystals for general purpose nonlinear
  optical frequency conversion}

\author{Ron Lifshitz}
\email{ronlif@tau.ac.il}
\affiliation{School of Physics and Astronomy, Raymond and Beverly
  Sackler Faculty of Exact Sciences, Tel Aviv University, Tel Aviv
  69978, Israel} 
\author{Ady Arie} 
\author{Alon Bahabad}
\affiliation{School of Electrical Engineering, Wolfson Faculty of
  Engineering, Tel Aviv University, Tel Aviv 69978, Israel}

\date{December 3, 2004}

\begin{abstract}
  We present a general method for the design of 2-dimensional
  nonlinear photonic quasicrystals that can be utilized for the
  simultaneous phase-matching of arbitrary optical
  frequency-conversion processes. The proposed scheme---based on the
  generalized dual-grid method that is used for constructing tiling
  models of quasicrystals---gives complete design flexibility,
  removing any constraints imposed by previous approaches. As an
  example we demonstrate the design of a color fan---a nonlinear
  photonic quasicrystal whose input is a single wave at frequency
  $\omega$ and whose output consists of the second, third, and fourth
  harmonics of $\omega$, each in a different spatial direction.
\end{abstract}


\maketitle

\begin{figure}[htbp]
  \centering
  \includegraphics[width=1.0\columnwidth]{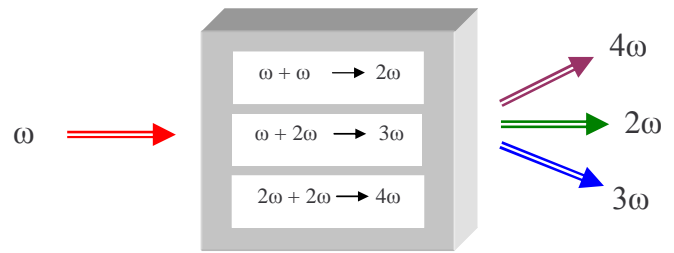}
  \caption{The color fan.}
  \label{fig:fan}
\end{figure}

The problem of {\it phase matching\/} in the interaction of light
waves in nonlinear dielectrics became immediately evident as the first
theories describing such interaction were developed~\cite{A62}.  Put
simply, nonlinear interaction is severely constrained in dispersive
materials because the interacting photons must conserve their total
energy and momentum.  Even the slightest wave-vector mismatch appears
as an oscillating phase that averages out the outgoing waves, hence
the term ``phase mismatch''. One approach for treating the problem
uses the birefringent properties of specific materials and by playing
with the polarizations of the interacting waves achieves phase
matching~\cite{G62,MTNS62}. A second approach, suggested over 4
decades ago~\cite{A62,BS70} and known today as
``quasi-phase-matching'', is to modulate the sign of the relevant
component(s) of the nonlinear dielectric tensor at the period of the
oscillating mismatched phase thereby undoing the averaging.
Quasi-phase-matching has been generalized from simple 1-dimensional
periodic modulation~\cite{FMJB92} to 2-dimensional periodic
modulation~\cite{B98,BRORH00,SK00,WG01,RPMS01} as well as
1-dimensional quasiperiodic
modulation~\cite{ZZM97,ZZQWGM97,FA99,FAUR02}, allowing greater
flexibility in phase-matching multiple frequency-conversion processes
within the same photonic crystal.  Here we present the full
generalization of the method that enables the design of nonlinear
photonic crystals that can simultaneously phase-match any arbitrary
set of frequency-conversion processes in any spatial direction. This
design flexibility is ideal for the realization of elaborate
multi-step cascading effects~\cite{KSS99,SSK03}, as demonstrated by
the color fan example (Fig.~\ref{fig:fan}) at the end of this article.

To understand how the method works it is convenient to adopt the view
taken in condensed matter systems. Recall that momentum conservation
is a direct consequence of having continuous translation symmetry. In
crystals, whether periodic or not, continuous translation symmetry is
broken, and momentum conservation is replaced by the less-restrictive
conservation law of crystal-momentum.  The total momentum of any set
of interacting particles in a crystal---whether they are electrons,
phonons, or photons---need only be conserved to within a wave vector
from the reciprocal lattice of the crystal, giving rise to so-called
{\it umklapp\/} processes.  Thus, all one needs to do is to construct
an artificial photonic crystal whose nonlinear susceptibility is
ordered on the appropriate length scale such that its reciprocal
lattice contains all the necessary wave vectors, needed to phase-match
any required frequency-conversion process. The linear susceptibility
should remain constant in space so that the dispersion relation will
not vary from point to point. We show here a method---borrowed from
the theory of quasicrystals~\cite{senbook}---to construct ordered
structures in real space with any given Fourier components. Such
structures are in general quasiperiodic, thus we are concerned here
with the design of nonlinear photonic quasicrystals (for a definition
of the term see the discussion in Ref.~\onlinecite{QCdef}).

To be more specific, consider for example, two incoming waves
$(\omega_1,\kv_1)$ and $(\omega_2,\kv_2)$ interacting in the plane of
a 2-dimensional dielectric whose total area is $A$, via its
lowest-order nonlinear susceptibility tensor $\chi^{(2)}$, to produce
an outgoing wave $(\omega_3,\kv_3)$, with $\omega_3=\omega_1 +
\omega_2$ and a wave vector mismatch $\dk = \kv_1 + \kv_2 - \kv_3$.
Assuming negligible depletion of the input waves, the amplitude of the
outgoing wave as it emerges from the interaction region $A$ is
\begin{equation}
  \label{eq:Eout}
  E_3(\dk) = \Gamma \int_A g(\rv) e^{i\dk\cdot\rv} d^2r,
\end{equation}
where $\Gamma$ is a constant that depends on the amplitudes of the
incoming waves and on the indices of refraction of all three waves,
and $g(\rv)$ is the relevant component of $\chi^{(2)}$, coupling the
three waves. Clearly, if $g$ is independent of $\rv$ the oscillating
mismatched phase averages out the output signal giving, for a
rectangular $A$, a two-dimensional {\it sinc\/} function that tends to
a delta function $\delta(\dk)$ as the interaction area $A$ tends to
infinity. In general, the output signal $E_3(\dk)$ is proportional to
the Fourier transform of $g(\rv)\varphi_A(\rv)$, where
$\varphi_A(\rv)$ is the characteristic function of the interaction
area $A$ (equal to 1 if $\rv\in A$, and 0 otherwise). We can design
this Fourier transform to peak at $\dk$ by ensuring that $\dk$ is a
vector in the reciprocal lattice of $g(\rv)$.

Consider more generally a set of $D$ nonlinear interactions of
triplets of waves, coupled by $\chi^{(2)}$ as in (\ref{eq:Eout}), with
$D$ 2-dimensional wave vectors $\k{j}$ [$j=1\ldots D$], describing
their mismatches. We wish to design a nonlinear photonic crystal
$g(\rv)$ that will simultaneously phase-match all $D$ interactions.
If, by accident, all $D$ mismatch vectors can be expressed as integral
linear combinations of {\it two\/} of them, $\k1$ and $\k2$, one can view
these two vectors as the basis of a 2-dimensional reciprocal lattice.
One then finds the dual real-space basis-vectors $\a1$ and $\a2$ whose
reciprocal vectors are $\k1$ and $\k2$ through the condition
\begin{equation}
  \label{eq:reciprocal}
  \a{i}\cdot\k{j} = 2\pi \delta_{ij},
\end{equation}
and generates a real-space lattice consisting of points at all
integral linear combinations of $\a1$ and $\a2$. The Fourier
transform of a set of delta functions at these lattice positions is a set of
delta function at all the reciprocal lattice positions generated by
integral linear combinations of the two mismatch wave vectors, as
required. This is the procedure that has been used until
now~\cite{B98,SK00,WG01,BRORH00}. 

In general, when the mismatch vectors do not all belong to a
reciprocal lattice of a periodic crystal, we need to construct a
quasicrystal whose Fourier transform contains all the mismatch
vectors. To do so we use a generalized version of de Bruijn's dual
grid method~\cite{dB81,GR86,RHM88,RHM89} used for the construction of
tiling models of quasicrystals. We begin by selecting a set of $D$
2-dimensional real-space vectors $\a{j}$ [$j=1\ldots D$], called
tiling vectors. We use a generalization of the orthogonality
condition~(\ref{eq:reciprocal}), known as the Ho condition~\cite{H86},
\begin{equation}
  \label{eq:Ho}
  \sum_{j=1}^D a_\mu^{(j)} \Delta k_\nu^{(j)} = 2\pi\delta_{\mu\nu}.
\end{equation}
Note that in the special case of a periodic structure the Ho
condition~(\ref{eq:Ho}) is simply the completeness relation that
follows from the orthogonality condition~(\ref{eq:reciprocal}) used to
define the reciprocal vectors. If we were now simply to generate all
points at integral linear combinations of the $D$ tiling vectors we
would get the unwanted outcome of a dense filling of real space. To
avoid this situation we construct the dual grid, whose topology
determines which of the integral linear combination of the tiling
vectors are to be included in the tiling.

To construct the dual grid we associate with each mismatch vector
$\k{j}=2\pi\nv^{(j)}/L_j$ an infinite family of parallel lines normal
to the unit vector $\n{j}$, separated by a distance $L_j$, and shifted
from the origin (in the direction of $-\n{j}$) by an amount $f_jL_j$,
for some chosen set of grid shifts $0\le f_j<1$.  This grid is dual to
the tiling in the sense that each {\it intersection\/} of lines in the
grid corresponds to a {\it tile\/} in the tiling whose edges are the
tiling vectors $\a{j}$ associated with the families of the
intersecting lines. In addition, each {\it cell\/} in the dual grid,
labeled by $D$ integers $n_j$, determines a {\it vertex\/} in the
tiling at position $\sum_{j} n_j \a{j}$, where $n_j$ is the number of
lines of the $j^{th}$ family separating the cell from the origin. Only
those linear combinations that correspond to cells in the dual grid
are included in the tiling.

The canonical choice of tiling vectors~\cite[section IV.C]{RHM88},
which is convenient for the analytical calculation of the Fourier
transform, is obtained by viewing the $D$ 2-dimensional mismatch
vectors as 2 $D$-dimensional vectors $(\Delta
k_\mu^{(1)},\ldots,\Delta k_\mu^{(D)})$ [$\mu=1,2$] spanning a
2-dimensional subspace of a $D$-dimensional vector space. One can then
choose $D-2$ additional $D$-vectors $(q_\mu^{(1)},\ldots,
q_\mu^{(D)})$ [$\mu=3\ldots D$], orthogonal to the first two, spanning
the remaining $(D-2)$-dimensional subspace.  This process adds $D-2$
components to each of the mismatch vectors extending them into the $D$
$D$-dimensional vectors $\K{j} = (\k{j},\q{j})$ that also span the
entire $D$-dimensional vector space. One can then find the dual basis
$\A{j}=(\a{j},\b{j})$, expanded into 2-dimensional tiling vectors
$\a{j}$ and $(D-2)$-dimensional extensions $\b{j}$, using a
$D$-dimensional orthogonality condition
\begin{equation}
  \label{eq:Dreciprocal}
  \A{i}\cdot\K{j} = 2\pi \delta_{ij},
\end{equation}
which implies a $D$-dimensional completeness relation
\begin{equation}
  \label{eq:Dcomplete}
  \sum_{j=1}^D A_\mu^{(j)} K_\nu^{(j)} = 2\pi\delta_{\mu\nu},
\end{equation}
thus satisfying the Ho condition~(\ref{eq:Ho}). Note that the choice
of the $D-2$ vectors $(q_\mu^{(1)},\ldots, q_\mu^{(D)})$ is not
unique, but once these are chosen the tiling vectors $\a{j}$ and their
$(D-2)$-dimensional extensions $\b{j}$ are uniquely determined by the
orthogonality condition~(\ref{eq:Dreciprocal}).

It can be shown~\cite[Eq. 5.5]{RHM88} that if $\rho(\rv)$ is a sum of
delta functions, at the positions determined by the dual grid method,
then its Fourier transform $\rho(\kv)$ is non-vanishing at most on the
lattice of integral linear combinations of the $\k{j}$. If the $\k{j}$
are linearly independent over the integers this immediately gives us
Bragg peaks at the positions required for multiple phase-matching. If
the mismatch vectors are integrally-dependent some of the Bragg peaks
may become extinguished. To avoid this situation one may simply choose
an independent subset of mismatch vectors to generate the structure,
whose Fourier transform will then contain all the required Bragg
peaks. Alternatively, one may refine the procedure for selecting the
tiling vectors. We shall consider the latter option in greater detail
elsewhere. The Fourier coefficient at $\kv$ is given by
\begin{equation}
  \label{eq:Fourier}
  \rho(\kv) = \frac1{v} \sum e^{i\qv\cdot\fv} \int_W {\text d}\sv\
  e^{i\qv\cdot\sv}, 
\end{equation}
where the sum---required only when the mismatch vectors are
linearly-dependent over the integers---is over all $\qv=\sum m_j\q{j}$
such that $\sum m_j\k{j} = \kv$, the shift vector $\fv=\sum f_j
\b{j}$, $W$ is the so-called $(D-2)$-dimensional window, given by the
set of all points $\sum \lambda_j \b{j}$ with $0\le \lambda_j <1$,
$\sv$ is a $(D-2)$-dimensional integration vector, and $v$ is the
volume of the primitive cell of the $D$-dimensional real-space lattice
generated by the vectors $\A{j}$.

Methods for modulating the nonlinear coefficient have been developed
for many $\chi^{(2)}$ materials, and in particular for
ferroelectrics~\cite{yamada93} and semiconductors~\cite{eyers01}.
Although typically the nonlinear coefficient is modulated only in one
direction, the modulation methods are based on planar technologies,
hence 2-dimensional modulation is possible~\cite{BRORH00}. In
ferroelectrics such as LiNbO$_3$ and KTiOPO$_4$, the modulation is
based on reversing the electrical domains in the $z$ direction of the
crystal, and thus provides either a positive or negative value of the
$\chi_{33}^{(2)}$ coefficient. In the following calculation, we have
assumed a minimum domain size of 2 $\mu m$, which can be achieved
using the electric field poling technique~\cite{FAUR02}. Thus the
actual nonlinear crystal $g(\rv)$ is extended over a finite area $A$,
and consists of positive domains of a given shape $S$, typically small
circles~\cite{SK00} or polygons such as hexagons~\cite{BRORH00} or
squares~\cite{RPMS01}, positioned at the vertices of the tiling in a
negative background. Using the convolution theorem, its actual Fourier
transform $g(\kv)$ is then
\begin{equation}
  \label{eq:ActualFourier}
  g(\kv) = \Delta\chi \left(\rho(\kv)\otimes
  \int_A e^{i\kv\cdot\rv} d^2r \right) \int_S e^{i\kv\cdot\rv} d^2r,
\end{equation}
where $\Delta\chi$ is the absolute difference between the positive and
negative values used for $\chi_{33}^{(2)}$, $\rho(\kv)$ is the sum
of delta functions whose amplitudes are given in (\ref{eq:Fourier}),
and $\otimes$ is the convolution operator. For example, if $S$ is a
circle of radius $R$ and $A$ a rectangle of sides $L_x\times L_y$, the
Fourier transform is
\begin{eqnarray}
  \label{eq:FourierExample}\nonumber
  g(\kv) &=& g(k_x,k_y) = 2SA\Delta\chi {J_1(kR)\over{kR}}\\
  &&\times\left(\rho(\kv)\otimes
  \left[\sinc\left({k_x L_x\over2}\right) 
  \sinc\left({k_y L_y\over2}\right)\right]\right)
\end{eqnarray}
where $k=|\kv|$, and $J_1$ is the first Bessel function. There is, of
course, much room for fine tuning of the design process---especially
in the choice of the domain shape $S$---but we have clearly
demonstrated that the required mismatch wave vectors indeed appear in
the Fourier transform of the nonlinear photonic quasicrystal. Note in
particular for the circle, that care must be taken in the choice of the
radius $R$ such that the first zero of the Bessel function, occuring at
$kR\simeq3.8317$, does not extinguish the required Bragg peaks.

\begin{table*}[tbp]
  \centering
  \begin{tabular}{|c|c|c|c|}\hline
  \text{Mismatch vector} & \text{Magnitude in} $(\mu m)^{-1}$ &
  \text{Direction in radians} & $g(\k{i})$ \\\hline
  $\k1=\dk_{11\rightarrow2}$ & $1.0183$ &  $0$      & 0.0404\\
  $\k2=\dk_{22\rightarrow4}$ & $2.8179$ &  $0.0456$ & 0.0431\\
  $\k3=\dk_{12\rightarrow3}$ & $0.9530$ & $-0.0966$ & 0.1671\\\hline
  \text{Tiling vector} & \text{Magnitude in} $\mu m$ &
  \text{Direction in radians}   &\\\hline
  $\a1$ & $10.330$  & $-1.4801$ &\\
  $\a2$ & $18.234$ &  $1.5001$ &\\
  $\a3$ & $42.297$ & $-1.5289$ &\\\hline
  \end{tabular}
  \caption{Calculated mismatch vectors for the color fan and the
  corresponding tiling vectors. The right-hand column lists the
  analytically-calculated Fourier coefficients for the three processes
  assuming an infinite structure and circular positive domains of
  radius $2.9212\mu m$ in a negative background. The corresponding values
  obtained in a 1-dimensional periodic crystal for first and third
  order processes are $2/\pi\simeq 0.6366$ and $2/3\pi\simeq
  0.2122$, respectively~\cite[p.~2635]{FMJB92}.  }
  \label{tab:vectors}
\end{table*}

\begin{figure}[htbp]
  \centering
  \includegraphics[width=0.8\columnwidth]{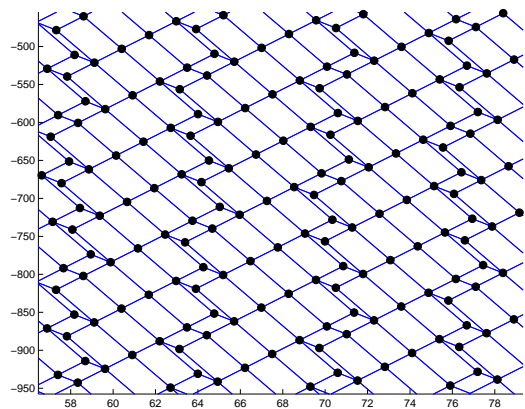}
  \caption{Structure of the photonic quasicrystal, produced with the
parameters of Table~\ref{tab:vectors}. Scale is in $\mu m$.}
  \label{fig:crystal}
\end{figure}

\begin{figure}[htbp]
  \centering
  \includegraphics[width=0.8\columnwidth]{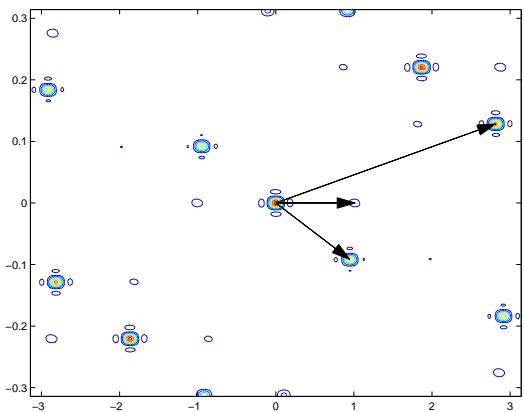}
  \caption{Numerical Fourier transform of the crystal in
    Fig.~\ref{fig:crystal} showing Bragg peaks at the positions of the
    required mismatch vectors, as indicated by arrows. Scale is in
    $(\mu m)^{-1}$.}
  \label{fig:fourier}
\end{figure}

As an example for our design procedure consider the color fan,
depicted in Fig.~\ref{fig:fan}, whose input is a fundamental beam at
frequency $\omega$ in the $\hat x$ direction and whose output consists
of the second, third, and fourth harmonics in different directions.
Note that the pair of interacting beams, in all three frequency
conversion processes, $\omega$ and $2\omega$, are travelling in the
same ($\hat x$) direction to allow for efficient interaction
throughout the full length of the photonic crystal. Using the
experimental values for the dispersion in lithium niobate at
$150^\circ$C, given in Ref.~\onlinecite{J97}, and taking an input beam
of wavelength $\lambda=1550nm$, the required mismatch vectors and the
corresponding tiling vectors, satisfying the Ho
condition~(\ref{eq:Ho}), are given in Table~\ref{tab:vectors}. In
order to keep all three mismatch vectors to be of the same magnitude
we use for the first process ($\omega+\omega\to2\omega$) a third order
quasi-phase-matching condition, as described for example in
Ref.~\onlinecite{FMJB92}. A section of the resulting photonic
quasicrystal, produced with these parameters using the dual grid
method, is shown in Fig.~\ref{fig:crystal}. Its Fourier transform is
shown in Fig.~\ref{fig:fourier}, clearly showing Bragg peaks exactly
at the required positions of the mismatch vectors.

The method described in this Letter enables to achieve efficient
frequency mixing of any arbitrary set of input waves, in order to
generate a chosen set of output waves with specified wavelengths and
propagation directions. We have shown the implementation of the
dual-grid method for a two-dimensional modulation of the nonlinear
coefficient. The method is, of course, applicable in any number of
dimensions. The implementation in one dimension is straightforward and
requires that the wave-vectors of the interacting waves, and thus the
mismatch wave-vectors, are all collinear. For example, the
one-dimensional Fibonacci structure~\cite{ZZM97} can easily be
generated by choosing two collinear mismatch wave-vectors having a
ratio of $\tau=(1+\sqrt5)/2$ between them.

An interesting device that is enabled by the proposed method is a
nonlinear mixer that can generate the same harmonic at several
different directions. This mixer can be highly useful for
cavity-enhanced nonlinear processes, such as resonant second harmonic
generation and optical parametric oscillation, with the unique feature
that all propagation directions in the cavity contribute to the
nonlinear process. In addition to the possibility to generate multiple
harmonics at multiple directions, as shown here, some interesting
applications of this method can benefit from the possibility of
generating large nonlinear phase shifts by cascaded nonlinear
processes. These applications include generation of multicolor optical
solitons, as well as all-optical deflection and switching of
light~\cite{SSK03}.   

\begin{acknowledgments}
  This research is supported by the Israel Science Foundation through
  Grant No.~278/00, and by the Israeli Ministry of Science. 
\end{acknowledgments}

\bibliography{gdmoptics}

\end{document}